\documentclass[preprint,twocolumn]{aastex62}
\usepackage{mathtools}

\graphicspath{{./}{figures/}}
\usepackage{lineno}
\usepackage{booktabs}
\usepackage{siunitx}
\usepackage{array}
\usepackage{xcolor}
\usepackage{colortbl}
\usepackage{ulem}

\shorttitle{S241125n}
\shortauthors{Zhang et al.}

\begin{document}
\title{LVK S241125n: Massive Binary Black Hole Merger Produces GRB in AGN Disk}

\correspondingauthor{}

\author[0000-0003-0368-384X]{Shu-Rui Zhang}
\affil{School of Astronomy and Space Science, University of Science and Technology of China, Hefei 230026, China}
\affil{Department of Astronomy, University of Science and Technology of China, Hefei, Anhui 230026, China}
\affil{Dip. di Fisica e Scienze della Terra, Universit\`a degli Studi di Ferrara, Via Saragat 1, I--44122 Ferrara, Italy}
\affil{ICRANet, P.zza della Repubblica 10, 65122 Pescara, Italy.}

\author[0000-0001-7959-3387]{Yu Wang}
\email{yu.wang@icranet.org}
\affil{ICRANet, P.zza della Repubblica 10, 65122 Pescara, Italy.}
\affil{ICRA - Dipartimento di Fisica, Sapienza Universit\`a di Roma, P.le Aldo Moro 5, 00185 Rome, Italy.}
\affil{INAF -- Osservatorio Astronomico d'Abruzzo, Via M. Maggini snc, I-64100, Teramo, Italy.}

\author[0000-0002-7330-4756]{Ye-Fei Yuan}
\email{yfyuan@ustc.edu.cn}
\affil{School of Astronomy and Space Science, University of Science and Technology of China, Hefei 230026, China}
\affil{Department of Astronomy, University of Science and Technology of China, Hefei, Anhui 230026, China}

\author[0000-0002-5674-0644]{Hiromichi Tagawa}
\affiliation{Shanghai Astronomical Observatory, Shanghai, 200030, China}

\author[0000-0002-9130-2586]{Yun-Feng Wei}
\affiliation{Institute of Fundamental Physics and Quantum Technology, Ningbo University, Ningbo, Zhejiang 315211, China}
\affiliation{School of Physical Science and Technology, Ningbo University, Ningbo, Zhejiang 315211, China}

\author[0000-0002-1343-3089]{Liang Li}
\affiliation{Institute of Fundamental Physics and Quantum Technology, Ningbo University, Ningbo, Zhejiang 315211, China}
\affiliation{School of Physical Science and Technology, Ningbo University, Ningbo, Zhejiang 315211, China}

\author[0000-0002-2242-1514]{Zheng-Yan Liu}
\affiliation{School of Astronomy and Space Science, University of Science and Technology of China, Hefei 230026, China}
\affiliation{Department of Astronomy, University of Science and Technology of China, Hefei, Anhui 230026, China}

\author[0000-0002-1330-2329]{Wen Zhao}
\affiliation{School of Astronomy and Space Science, University of Science and Technology of China, Hefei 230026, China}
\affiliation{Department of Astronomy, University of Science and Technology of China, Hefei, Anhui 230026, China}

\author[0000-0002-3539-7103]{Rong-Gen Cai}
\affiliation{Institute of Fundamental Physics and Quantum Technology, Ningbo University, Ningbo, Zhejiang 315211, China}

\begin{abstract} 
Recently, the gravitational-wave (GW) event S241125n, detected by LIGO/Virgo/KAGRA (LVK), has been reported to coincide with a candidate detected by Swift-BAT/GUANO and an X-ray candidate found by FXT onboard of Einstein Probe (EP) and confirmed by Swift-XRT. We estimate that the joint false alarm rate (FAR) for the three candidates is 1 / 30 yr and that the corresponding false alarm probability (FAP) is $\mathrm{FAP}_{\rm triple} = 0.037$ ($1.8 \sigma$). The coincidence between the GW and GRB could be an interesting test of their origin and open attractive opportunities for multi-messenger observations, if they are actually associated. Motivated by this, we propose a theoretical model in which a binary black hole (BBH) merger occurs within an active galactic nucleus (AGN) disk. The typically massive and significantly kicked merger remnant accretes disk material at hyper-Eddington rates, and the resulting jet could lead to the GRB associated with the GW event. As the jet interacts with the gas in the AGN disk, the shock breakout produces a Comptonized spectrum, consistent with an unusually soft photon index of the GRB prompt emission observed by Swift-BAT following LVK S241125n. Meanwhile, strong absorption and dust extinction of the afterglow by the high column density typical of AGN disks could explain the unusually hard spectrum observed in the X-ray band by EP, as well as the non-detection of an optical counterpart. Our model is predictive, and we highlight the importance of further constraining the orbital eccentricity of the merger and conducting deep-field observations of the host galaxy to test our explanation.
\end{abstract}

\section{Introduction} 
\label{sec:intro}
Typically, short gamma-ray bursts (GRBs) are believed to originate from the merger of binary systems containing at least one neutron star \citep{2014ARA&A..52...43B}. Binary black hole (BBH) mergers are generally not expected to produce GRBs. However, there are exceptions where GRB production becomes possible, such as when the black holes (BHs) are charged \citep{2019ApJ...873L...9Z}, when the BH binary forms from two clumps in a dumbbell configuration \citep{2016ApJ...819L..21L}, when the merger of two BHs occurred within the envelope of a massive star that gave birth to one of the BHs \citep{2016ApJ...821L..18P}, and when they are located in high-density gas environments like active galactic nucleus (AGN) disks \citep{2023ApJ...950...13T}.

In AGN disk environments, as natural astrophysical sites for the activities of compact objects, BH binaries could form from stellar binaries born within the disk \citep{2025MNRAS.537.3396E}, tied by a circumbinary disk \citep{2022ApJ...928L..19L,2022MNRAS.517.1602L,2023MNRAS.522.1881L,2024MNRAS.529..348L}. They can also form from initially isolated BHs that are captured into binary systems through dynamical interactions \citep{2022Natur.603..237S,2022ApJ...934..154L,2023MNRAS.518.5653B} or gas interactions \citep{2023MNRAS.523.1126D,2023MNRAS.521..866R,2024arXiv241212086R,2024MNRAS.531.4656W,2024ApJ...962..143Q,2024ApJ...972..193D}. Once a binary BH merges within an AGN disk, GWs are produced, typically retaining information about residual orbital eccentricity \citep{2021ApJ...907L..20T,2024MNRAS.534L..58R}. Meanwhile, the remnant BH accretes surrounding material, often at hyper-Eddington rates \citep{2020PhRvL.124y1102G,2021ApJ...916L..17W,2021ApJ...916..111K}. The kick velocity resulting from the merger further enhances both the accretion rate and the duty cycle \citep{2024ApJ...961..206C,2023ApJ...950...13T}. During hyper-Eddington accretion within the disk, the BH is typically surrounded by an outflow-dominated circum-BH disk \citep{2021ApJ...911L..14W,2022ApJ...927...41T,2023ApJ...948..136C}. Magnetic fields surrounding the BH may accumulate \citep{2011ApJ...737...94C} and be further amplified by the outflow \citep[e.g.,][]{2020MNRAS.494.3656L}. Simultaneously, accretion could further increase the BH's spin residual from the merger, although turbulence within the AGN disk might prevent it from spinning up significantly \citep{2023MNRAS.522..319C}. 
Additionally, the BH spin magnitude is significantly amplified during the merger \citep{2005PhRvL..95l1101P}. 
Under these conditions, jets are expected to be launched via the Blandford–Znajek (BZ) mechanism \citep{blandford1977electromagnetic}.

The jet driven by accretion interacts with the disk material, creating shocks. The diffusion of photons is slower than the propagation of the shock, causing photons to be initially trapped \citep{2021ApJ...911L..19Z,2022ApJ...932...80Y}. Once the shock propagates to a height where photon diffusion becomes faster than shock propagation, photons escape, enabling the production of GRBs \citep{2010ApJ...725..904N,2012ApJ...747...88N}. Such GRBs are theoretically predicted to differ from standard ones in both their burst duration and the spectral energy distribution (SED) of 
the prompt \citep{2023ApJ...950...13T} and 
afterglow emissions \citep{2022ApJ...938L..18L,2022MNRAS.516.5935W}, 
as well as in possible bright shock cooling emission \citep{2024ApJ...966...21T}. 
They may even be choked, depending on the density and scale height of the disk \citep{2021ApJ...911L..19Z, zhang2024propagation}. Nevertheless, certain notable observational effects—especially the absorption and extinction of the afterglow caused by the high column density characteristic of AGN disks—have not yet been thoroughly investigated.

Gravitational wave (GW) events associated with GRBs are valuable multi-messenger sources. They can, in principle, be used to explore the mechanisms of GRB production, constrain the hosts of GW sources \citep{2022MNRAS.514.2092V,veronesi2023most,2025MNRAS.536..375V}, and refine cosmological parameters since the redshift and luminosity distance of the source can be determined separately through GRB and GW observations \citep{2020arXiv200914199M,2021ApJ...908L..34G,2024MNRAS.531.3679A,2024PhRvD.110h3005B,2024PhRvL.133z1001M}. Furthermore, 
it provides an opportunity to study the disk's physical properties if it occurs in an AGN disk.
However, observationally, only a handful of GRBs that may be associated with binary mergers in AGN disks have been identified (i.e., GRB 191019 \citep{2023NatAs...7..976L,2023ApJ...950L..20L} and GW 150914-GBM \citep{2016ApJ...826L...6C}), and even fewer have been linked to GWs \citep{2016A&A...593L..10B}, with the association still being controversial (i.e., GW 150914-GBM, \citealt{2016ApJ...827L..38G,2018ApJ...853L...9C,2023ApJ...950...13T}). Additionally, the recent GW event S241125n, detected by LIGO/Virgo/KAGRA (LVK), has been reported to coincide with a short GRB and X-ray afterglow emission.

In this paper we investigate a BBH merger occurring within an AGN disk, focusing on the resulting GRB in both its prompt and afterglow signatures. The new source S241125n could be an interesting test the model if the GW and GRB are actually associated.
The structure of this paper is as follows:
In Section \ref{sec:2}, we summarize the GW signal of S241125n together with the multi-wavelength results from X-ray and gamma-ray detectors. Section \ref{sec:3} explores the mechanisms for binary BH mergers producing GRBs in AGN disks, the associated prompt emergence and afterglow absorption features, and model fitting for S241125n. Section \ref{sec:4} discusses current limitations and alternative origins. Finally, Section \ref{sec:5} presents the conclusions.

\section{Observation}
\label{sec:2}

\begin{figure}
\centering
\includegraphics[width=1\linewidth]{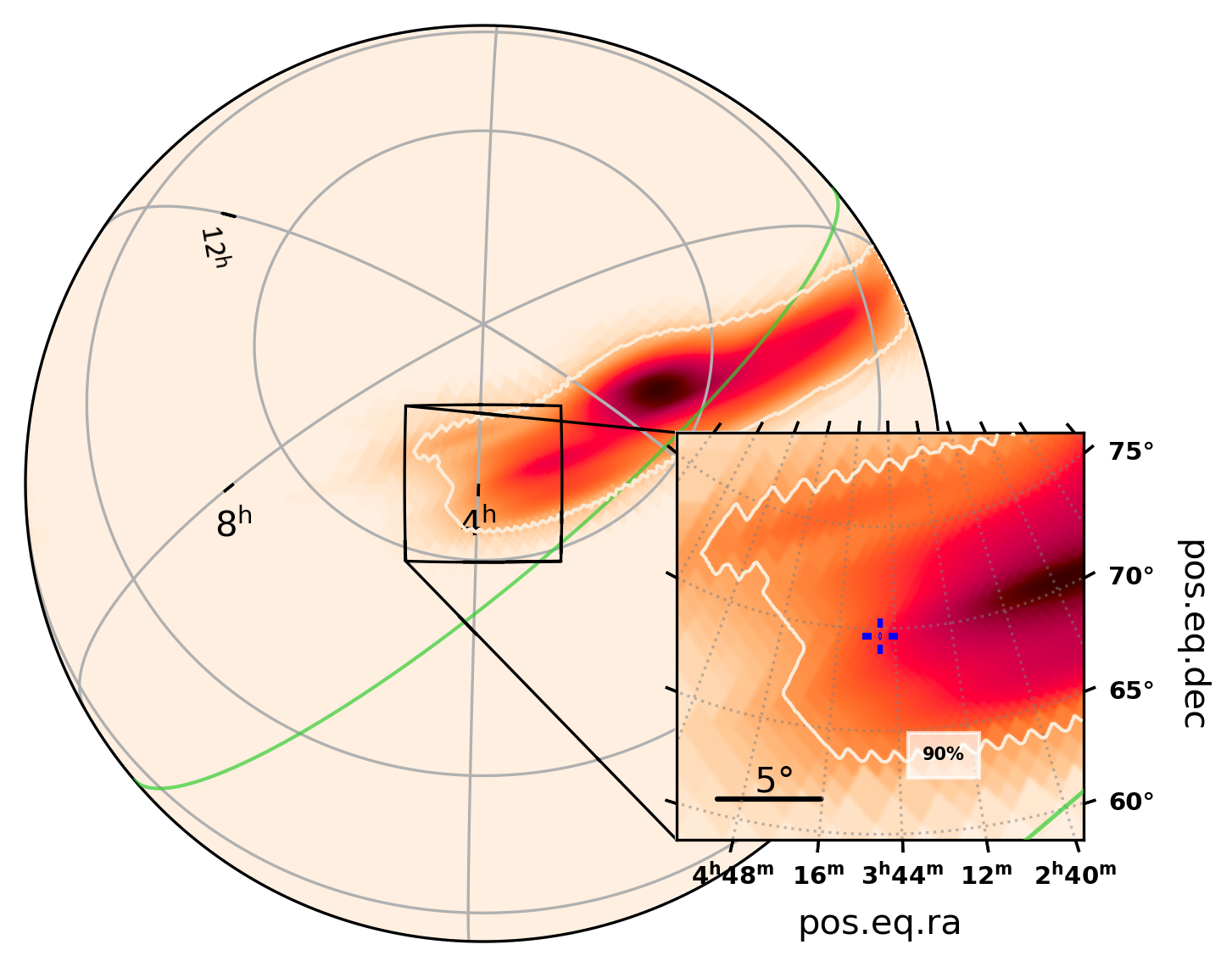} 
\caption{\textbf{GW skymap} of LVK S241125n and the GRB location. The color bar of the skymap represents the relative probability density of the GW source location. The white solid line represents the $90\%$ confidence level contour, while the blue cross in the inset indicates the position of the candidate electromagnetic counterparts. The green curve indicates the Galactic plane.} 
\label{fig:GWmap}
\end{figure}
\textbf{Gravitational Wave:} The compact binary merger candidate LIGO/Virgo/KAGRA (LVK) S241125n was identified during real-time data processing by the LIGO Hanford Observatory (H1), LIGO Livingston Observatory (L1), and Virgo Observatory (V1) on 2024 November 25 at 01:01:16.780 UTC ($T_0$) from a posteriori luminosity distance of $d_{\rm L} = 4173 \pm 1590$ Mpc (redshift $z=0.73$) \citep{2024GCN.38305....1L}. The GW signal was classified with probabilities as binary BH (BBH; $>99\%$), terrestrial ($<1\%$), binary neutron star (BNS; $<1\%$), or neutron star-BH (NSBH; $<1\%$). Noise transients (glitches) detected in the LIGO Hanford data may affect the signal's parameters or its statistical significance \citep{2024GCN.38305....1L}. 

Figure~\ref{fig:GWmap} shows the GW skymap of LVK S241125n, generated by the Bilby pipeline \citep{ashton2019bilby} and plotted using the \texttt{Bilby.offline0.multiorder.fits} file\footnote{\url{https://gracedb.ligo.org/superevents/S241125n/view/}}. The $90\%$ credible region covers an area of 2196~deg$^2$, which occupies $f_\mathrm{sky} = 5.32\%$ of the whole sky; the candidate electromagnetic counterparts lie within it.

We estimate that the joint false alarm rate (FAR) of the three candidates (GW + prompt and afterglow of the GRB-like event) is $\mathrm{FAR}_{\rm triple} = 1.06 \times 10^{-9}~\mathrm{Hz}\ (1 / 30 \rm yr)$; see details in Appendix B.
This provides a statistical analysis of the potential association between the GW and GRB events of S241125n, demonstrating that their physical connection is favored.

\begin{figure}
\centering
\includegraphics[width=1\linewidth]{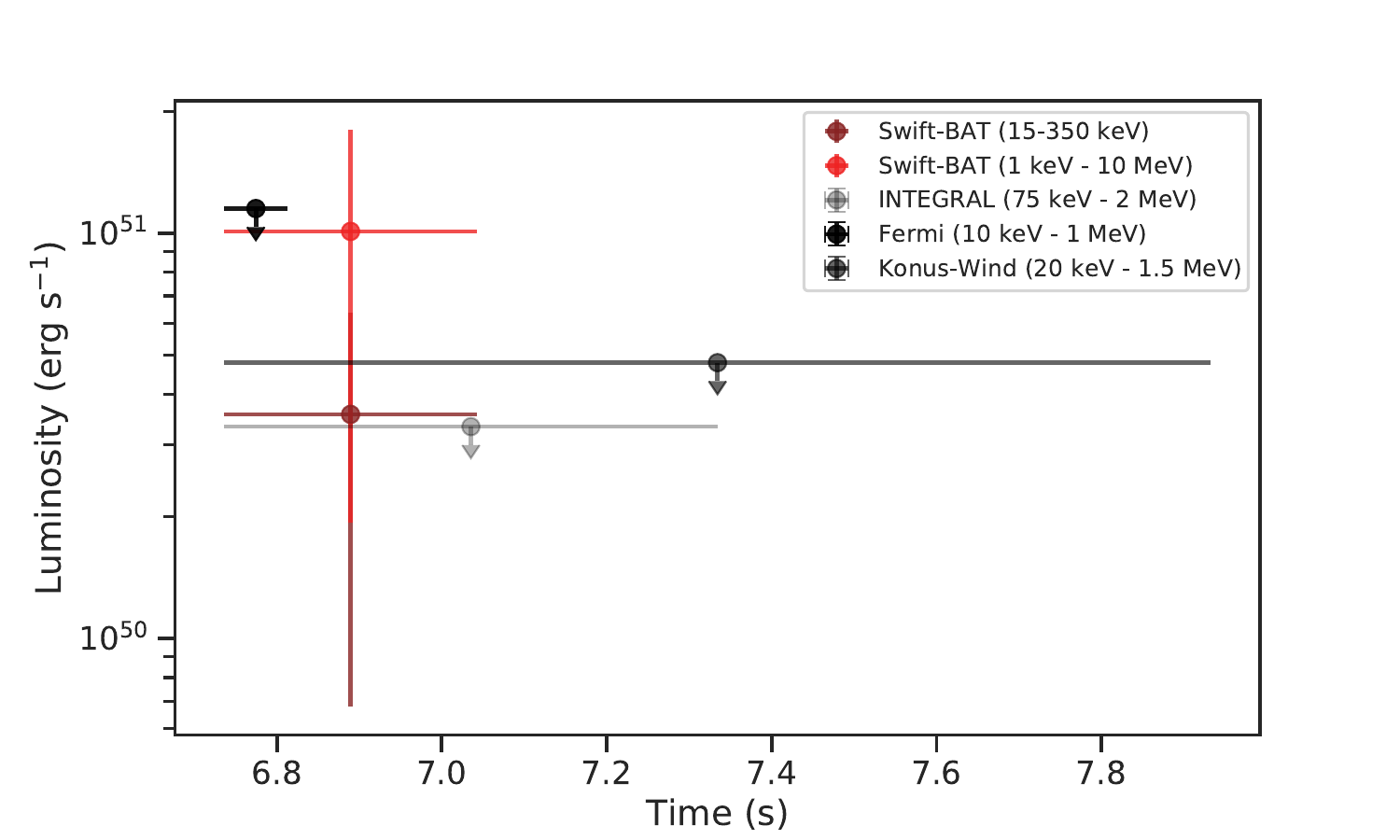} 
\caption{\textbf{Prompt emission luminosity} estimated by Swift, INTEGRAL, Fermi and Konus Wind satellites. The GW trigger time as the event start time ($T_0$) and time is in the rest-frame.}
\label{fig:prompt-emission}
\end{figure}

\begin{figure}
\centering
\includegraphics[width=1\linewidth]{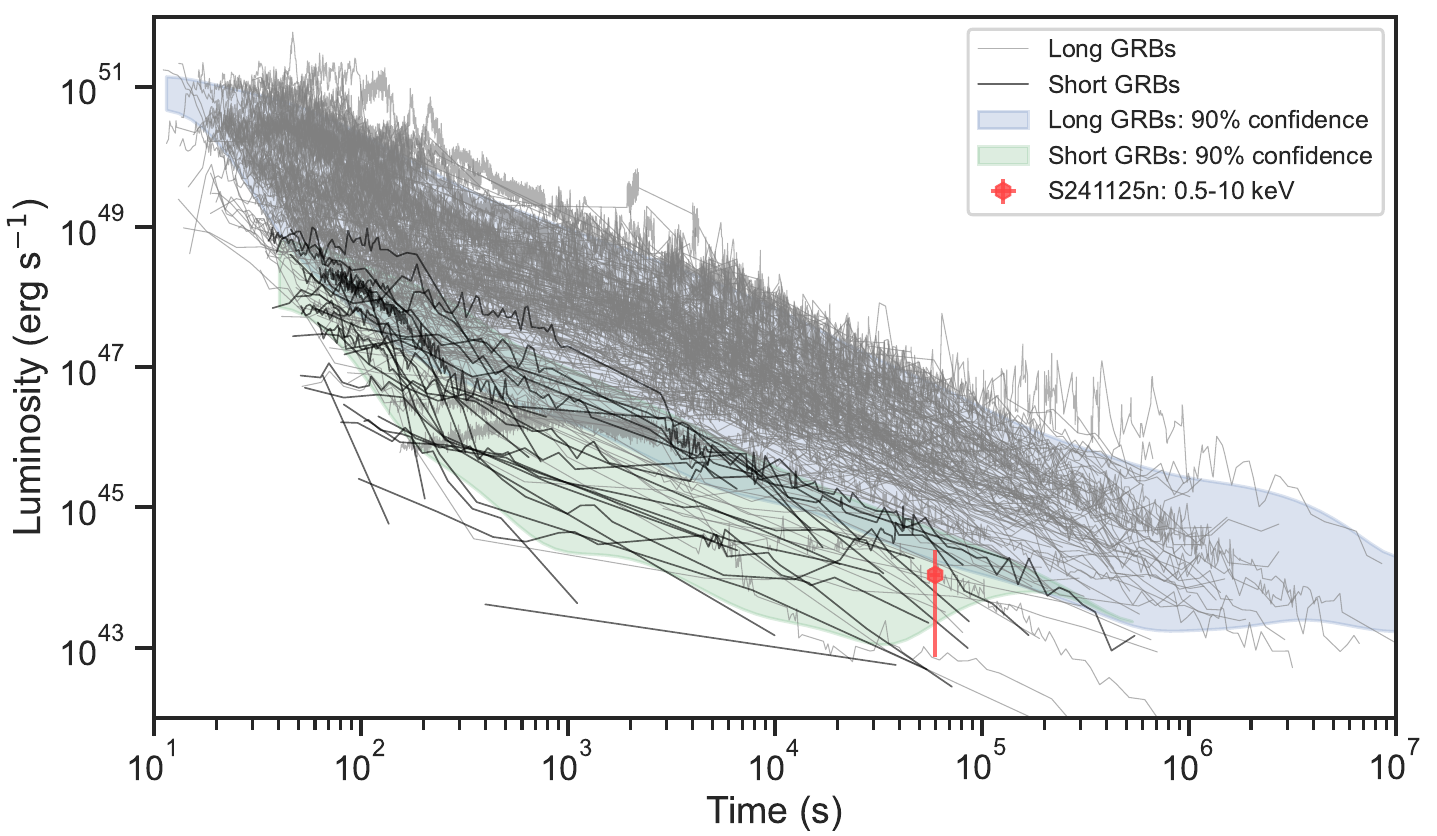} 
\caption{\textbf{The X-ray afterglow} data point of S241125n (in red) is obtained using the observation of Einstein Probe within the energy range of 0.5-10 keV. For comparison, background data from Swift-XRT observations include 217 long GRBs (in gray) and 31 short GRBs (in black), covering 0.3-10 keV. The shaded regions correspond to the 90\% of the confidence region generated from the data directly. It can be observed that S241125n is consistent with the luminosity of short GRBs. Time is in the rest-frame.}
\label{fig:afterglow}
\end{figure}

\begin{figure*}
\centering
\includegraphics[width=0.85\linewidth]{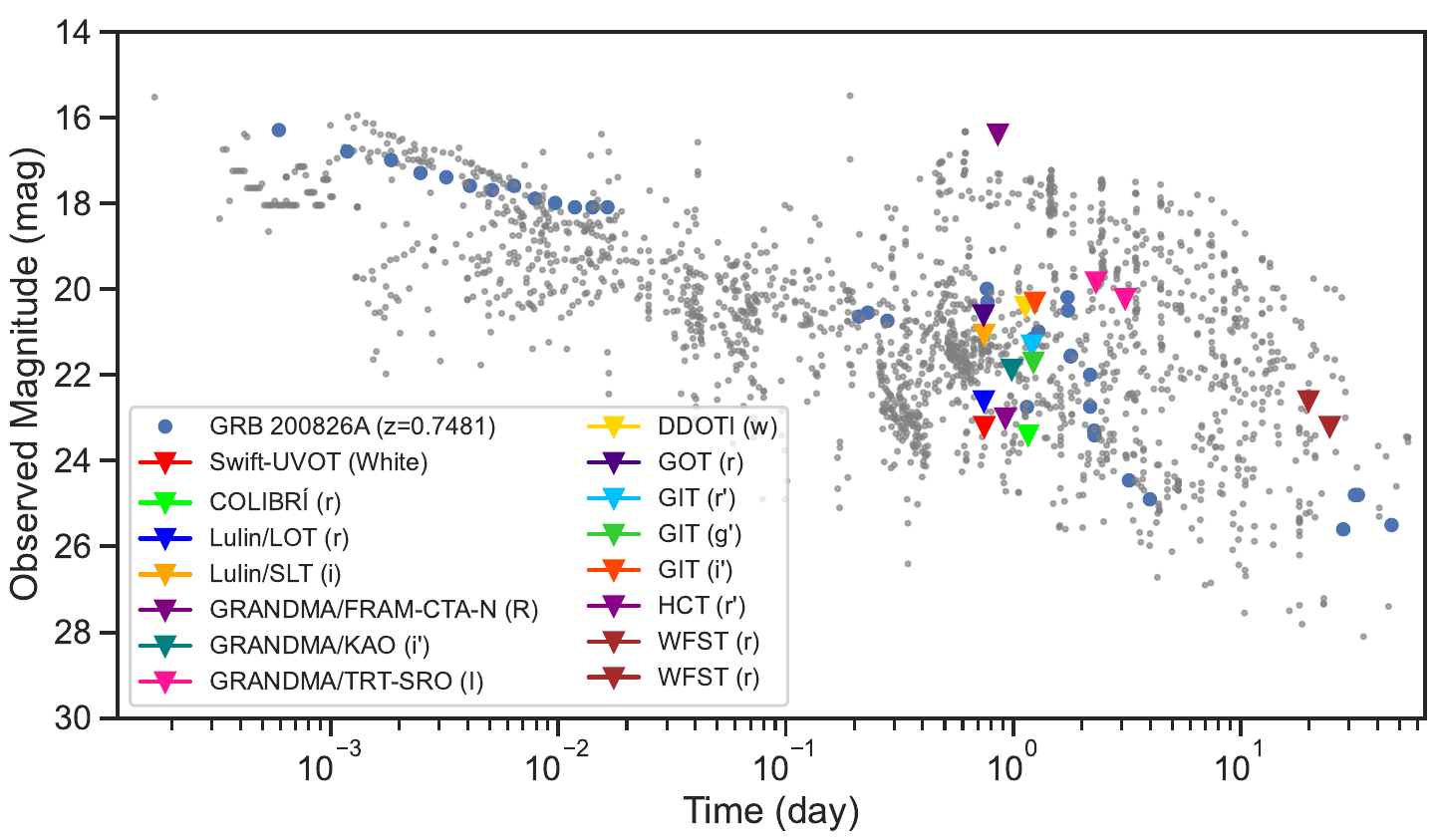} 
\includegraphics[width=0.9\linewidth]{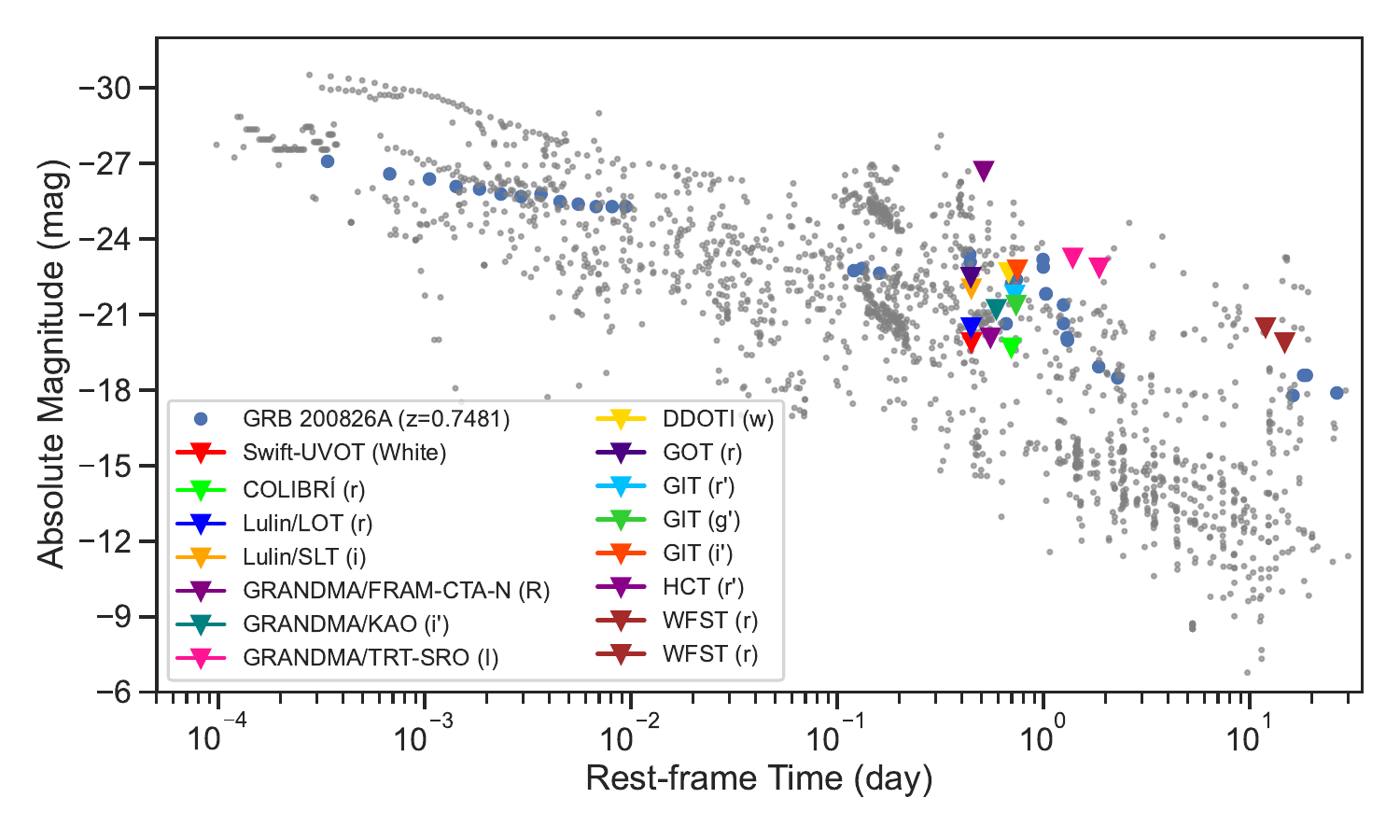}
\caption{\textbf{The optical observations of S241125n} include upper limits provided by Swift, COLIBRÍ, Lulin, GRANDMA, DDOTI, GOT, GIT and HCT telescopes.\textbf{Upper panel} shows the observed magnitude, and the \textbf{lower panel} shows the absolute magnitude. For comparison, the background data consists of optical observations covering multi-bands from 29 short GRBs (in gray) from \citet{2024MNRAS.533.4023D}. Among them, GRB 200826A (in blue), which has a similar redshift to S241125n, is highlighted. Time and magnitude are in the observer's frame. It should be noted that the reference background magnitudes have already been corrected for Galactic extinction, whereas the data for S241125n are mainly the raw observations directly taken from the GRB Coordinates Network (GCN). According to the Milky Way extinction in the direction of S241125n,  $E(B-V) = 0.4678$  mag \citep{2011ApJ...737..103S}, the extinction-corrected values for S241125n will be approximately 0.5 to 2 mag brighter than those shown in the Figure, for examples, $A_{i'} = 0.927$ mag, $A_{r'} = 1.224$ mag and $A_{g'} = 1.739$ mag.}
\label{fig:optical-limits}
\end{figure*}


\vspace{0.5em}
\textbf{Gamma-ray Counterpart:} Following the detection of LVK S241125n, the Swift Burst Alert Telescope (Swift-BAT) followed the notice sent by the Gamma-ray Urgent Archiver for Novel Opportunities \citep[GUANO,][]{2020ApJ...900...35T} and identified a candidate gamma-ray counterpart of trigger ID 754189311 with a square root of the test statistic of 7.41, starting at $T_0$+11.264 s and analyzed in a 0.512 s time bin (GCN Circulars 38308). The BAT position of the candidate was reported as RA, Dec = $58.079^\circ, +69.689^\circ$ with an estimated uncertainty of 5 arcmin (50\% containment) \citep{2024GCN.38308....1D}. 
Spectral analysis of the gamma-ray counterpart revealed a Comptonized spectrum (cutoff power law) with $E_{\text{peak}} = 49 \, \text{keV}$, photon index $= -2.2$, and flux in the 15-350 keV range of $f=(1.1_{-0.3}^{+0.2}) \times 10^{-7} \, \text{erg} \, \text{cm}^{-2} \, \text{s}^{-1}$ \citep{2024GCN.38351....1D}. While this is the case, still keep in mind that the spectral index is unconstrained in the range of -1.4 to -3.0, and the allowed values of $E_{\text{peak}}$ vary with it \citep{2024GCN.38351....1D}. 

We then calculated the isotropic-equivalent luminosity ($L_{\gamma,\text{iso}}$) of the gamma-ray counterpart using the observed flux and the luminosity distance derived from the
GW signal. Using the standard relation $L = 4 \pi k d_L^2 f$ where $k$ is the $k$-correction accounting for the cosmological redshift \citep{2001AJ....121.2879B}, we computed the rest-frame luminosity in the 15-350 keV range to be $(3.57^{+2.80}_{-2.89}) \times 10^{50} \, \text{erg} \, \text{s}^{-1}$, with a $k$-correction factor of $1.56$. To estimate the luminosity in the broader 1 keV--10 MeV range, we extrapolated the Comptonized spectral model, obtaining $(1.01^{+0.79}_{-0.82}) \times 10^{51} \, \text{erg} \, \text{s}^{-1}$ with a $k$-correction factor of $4.40$. The uncertainties reflect contributions from both the flux measurement and the luminosity distance. For a burst of observed duration $\Delta t = 0.512 \, \text{s}$, the isotropic-equivalent energy ($E_{\gamma,\text{iso}}$) in the 15–350 keV range is calculated as $E_{\gamma,\text{iso}} = L_{\gamma,\text{iso}} \times \Delta t / (1+z)$, resulting in $(1.09^{+0.86}_{-0.88}) \times 10^{50} \, \text{erg}$. Similarly, the energy in the 1 keV-10 MeV range is $(3.09^{+2.42}_{-2.50}) \times 10^{50} \, \text{erg}$.

No gamma-ray counterpart was detected by INTEGRAL \citep{2024GCN.38311....1S}, Fermi \citep{2024GCN.38316....1S}, and Konus-Wind \citep{2024GCN.38321....1R}, and only upper limits were computed. INTEGRAL reported a 3-sigma upper limit for the $75-2000$ keV range of $1.6 \times 10^{-7} \text{erg} ~\text{cm}^{-2}$, assuming a burst duration of less than 1 second and adopting a typical short GRB spectrum \citep{2024GCN.38311....1S}, corresponding to the luminosity upper limit of $3.33 \times 10^{50}~ \text{erg}~ \text{s}^{-1}$. Similarly, Fermi, using a hard-spectrum template typical for short GRBs, placed a $3$-sigma upper limit on the flux in the $10-1000$ keV range of $5.5 \times 10^{-7}~\text{erg}~ \text{s}^{-1}\text{cm}^{-2}$ over a $0.128$-second timescale \citep{2024GCN.38316....1S}, corresponding to the luminosity of $1.15 \times 10^{51} \text{erg}\ \text{s}^{-1}$. Konus-Wind estimated a $90\%$ confidence upper limit on the flux in the $20-1500$ keV range of $2.3 \times 10^{-7}~ \text{erg}~\text{s}^{-1} \text{cm}^{-2}$ for a burst with a spectrum similar to that of GRB 170817A \citep{2024GCN.38321....1R}, corresponding to the luminosity $4.79 \times 10^{50} ~\text{erg}~ \text{s}^{-1}$. $k$-correction has not been applied on computing the upper limits of luminosities.

Figure \ref{fig:prompt-emission} compiles the observational data of prompt emission from above telescopes. It can be seen that the isotropic luminosity falls within the range of $10^{50} - 10^{51} \, \text{erg}~\text{s}^{-1}$, which is consistent with the typical luminosity of short GRBs. Based on BAT data, the isotropic gamma-ray energy $E_{\gamma,\text{iso}} = (3.09^{+2.42}_{-2.50}) \times 10^{50} \, \text{erg}$ and the rest-frame peak energy is calculated as $ E_{p,\text{rest}} = 49 \times (1+z) = 85~\text{keV}$, which aligns with the empirical relation between $E_{p,\text{rest}}$ and $E_{\gamma,\text{iso}}$ for short GRBs \citep{2017SSRv..207...63W}. Appendix A presents a detailed test demonstrating that the GRB is intrinsically short. However, the photon index of this burst reaches $-2.2^{+0.6}_{-0.8}$, which is most likely 
softer than the usual low-energy spectral index of approximately $-1.5$ for GRB prompt emissions. This may indicate a special radiation mechanism for the prompt emission or a different propagation effect of the radiation.

\vspace{0.5em}
\textbf{X-Ray Candidates:} The Swift XRT performed follow-up observations spanning from $T_0$+55 ks to $T_0$+74 ks after the GW trigger \citep{2024GCN.38324....1P}. 5 uncatalogued X-ray sources \citep{2024GCN.38324....1P} then lately updated to 15 uncatalogued X-ray sources were detected\footnote{\url{https://www.swift.ac.uk/LVC/S241125n/}}. However, none of these sources were significantly bright to be the confident counterparts of the GW signal, and none of them are confirmed to be associated with a galaxy. 

Einstein Probe (EP) conducted follow-up observations at $T_0$+94 ks with an exposure of $\sim$11 ks \citep{2024GCN.38345....1W}. Within the 5-arcmin BAT error circle, one X-ray source was detected by both EP modules at RA, Dec = $58.1097^\circ, +69.6392^\circ$ with a positional uncertainty of 10 arcsec (90\% confidence). Spectral analysis in the $0.5-10$ keV band revealed an absorbed power-law model with a photon index of $0.43^{+0.76}_{-0.74}$ and a flux of $(1.17^{+1.18}_{-0.63}) \times 10^{-13} \, \text{erg} \, \text{s}^{-1} \, \text{cm}^{-2}$ \citep{2024GCN.38345....1W}. Considering the $k$-correction of $k=0.45$, the corresponding luminosity is $(1.09^{+1.37}_{-1.01}) \times 10^{44} \, \text{erg} \, \text{s}^{-1}$.

The currently available public data have not confirmed an associated X-ray transient. Here, we assume that the EP data originate from LVK S241125n, i.e., interpreting as the X-ray afterglow of the corresponding GRB (see Appendix B for more details on the association assessment). From Figure \ref{fig:afterglow}, it can be seen that the EP data point falls within the typical short GRB afterglow region.The photon index observed by EP is $0.43^{+0.76}_{-0.74}$, which is significantly softer than the typical afterglow phase index of $-2$. This may suggest a different radiation mechanism or a different surrounding environment, such as a high hydrogen density of the host galaxy, which could result in strong absorption of low-energy X-rays.

\vspace{0.5em}
\textbf{Optical Observations:} Despite extensive follow-up efforts across multiple observatories, no definitive optical counterpart to the GW event LVK S241125n has been identified and only upper limits are obtained, including Lulin Observatory, COLIBRÍ telescope, Himalayan Chandra Telescope (HCT), GROWTH-India telescope, Gaoyazi/GOT telescope, DDOTI/OAN wide-field imager, MMT  telescope, Swift-UVOT and GRANDMA network \citep{2024GCN.38314....1C, 2024GCN.38317....1W, 2024GCN.38322....1S, 2024GCN.38325....1M, 2024GCN.38328....1J, 2024GCN.38329....1B,2024GCN.38333....1R, 2024GCN.38334....1A, 2024GCN.38350....1K,2024GCN.38396....1A}, summarized in Figure \ref{fig:optical-limits}.  

Most of the optical observations were conducted around one day after the event, with no early data available before 0.5 days. We additionally utilized Wide Field Survey Telescope (WFST; \cite{WangTG_2023}) in r-band to monitor the candidates reported by EP \citep{2024GCN.38345....1W} in late time (2024-12-14T22:17 and 2024-12-19T16:51 UTC). No significant variability was detected within the EP error circle, with 5$\sigma$ limiting magnitudes of 22.5-23.5 mag (see Figure 4), compared with Pan-STARRS1 archival images \citep{Finkbeiner_2016,Flewelling_2020}. The existing upper limits fall within the reasonable range for short GRBs \citep{2024MNRAS.533.4023D}, and it has not yet been confirmed whether a transient optical source exists. The data from MMT \citep{2024GCN.38333....1R} are not included in Figure \ref{fig:optical-limits}. MMT specifically observed the error circle of S241125n\_X2 \citep[the second X-ray source detected by Swift-XRT within the BAT error region,][]{2024GCN.38324....1P} and identified five optical sources with magnitudes ranging from 25.4 to 18.9, it has not been confirmed whether one of these sources is associated with LVK S241125n.

Based on the above data and the reference GRB 200826A, which has a similar redshift, the optical magnitude of S241125n is very likely to be fainter than 24 mag after a few days. Additionally, this GRB is located at a Galactic latitude of approximately $12$ degrees, near the Galactic plane ($<10$ degrees; see Figure \ref{fig:GWmap}), which increases the observational challenges.  Observing such a faint source requires large ground-based telescopes or space telescopes. If the host galaxy of this GRB experiences significant extinction as it could be, space-based infrared telescopes such as the James Webb Space Telescope \citep[JWST,][]{2006SSRv..123..485G} would be necessary for follow-up observations.

\vspace{0.5em}
\textbf{Light-curve Comparison:} 
The luminosity evolution of S241125n can be compared with the general distributions of short gamma-ray bursts. The prompt isotropic luminosity derived from \textit{Swift}-BAT lies in the range $10^{50}$--$10^{51}\,\mathrm{erg\,s^{-1}}$, consistent with the typical luminosity band of short bursts \citep{2010ApJ...716.1178A,2016ApJ...831..178R}. The isotropic-equivalent energy $E_{\gamma,\mathrm{iso}} \simeq 3\times10^{50}\,\mathrm{erg}$ and the rest-frame peak energy $E_{\mathrm{p,rest}}\simeq 85\,\mathrm{keV}$ also fall within the broader $E_{\gamma,\mathrm{iso}}-E_{\mathrm{p,rest}}$ relation of short GRBs \citep{2013MNRAS.430..163Q,2021NatAs...5..911Z}.

In the X-ray band, the Einstein Probe detection in the $0.5$--$10\,\mathrm{keV}$ range aligns with the luminosity distribution of short GRB afterglows shown in Fig.~\ref{fig:afterglow}. The inferred luminosity of order $10^{44}\,\mathrm{erg\,s^{-1}}$ lies within the spread of short GRBs at comparable rest-frame times, although the measured photon index is harder than the canonical afterglow value.

Fig.~\ref{fig:optical-limits} includes upper limits from multiple facilities of S241125n, and the distribution of optical afterglows from 29 short GRBs. Among these, GRB~200826A is highlighted because it lies at a similar redshift ($z=0.7481$) and shows similar brightness. This comparison indicates that S241125n falls within the known diversity of short GRB afterglows.

The first (binary BH merger) gravitational-wave event, GW150914 \citep{2016PhRvL.116f1102A}, has also been proposed to have an associated bright short GRB detected by the Fermi Gamma-ray Burst Monitor (GBM), GW150914-GBM \citep{2016ApJ...826L...6C}, with $L \sim 10^{49}$ erg s$^{-1}$, 0.4 s after the GW event, while no obvious afterglow candidate has been observed \citep{2016PASJ...68L...9M}. Additionally, the very heavy BH merger event GW190521 \citep{2020PhRvL.125j1102A} has been found to have spatial coincidence with a peculiar optical flare ZTF 19abanrhr (with luminosity $L \sim 10^{45}$ erg s$^{-1}$, duration $\sim 1$ month, and a 18-day temporal association) in the AGN J124942.3+344929 \citep{2020PhRvL.124y1102G}, while no corresponding explicit high-energy emission has been confirmed. Thus, LVK S241125n is a known rare GW candidate from a binary BH merger, detected with a putative coincident short GRB with prompt emission and afterglow (again, see Appendix B for more details on their association possibility analysis).

We can summarize the observations that the energy and luminosity (under the associated assumption \( d_L(\mathrm{GW}) = d_L(\mathrm{GRB}) \)) of S241125n are similar to those of short GRBs, but the observed spectral characteristics differ.

\section{BH Binary Merger in an AGN Disk as a Possible Scenario}
\label{sec:3}

\subsection{Prompt Emergence}
\label{sec:3.1}

\begin{figure*}
\centering
\includegraphics[width=1\linewidth]{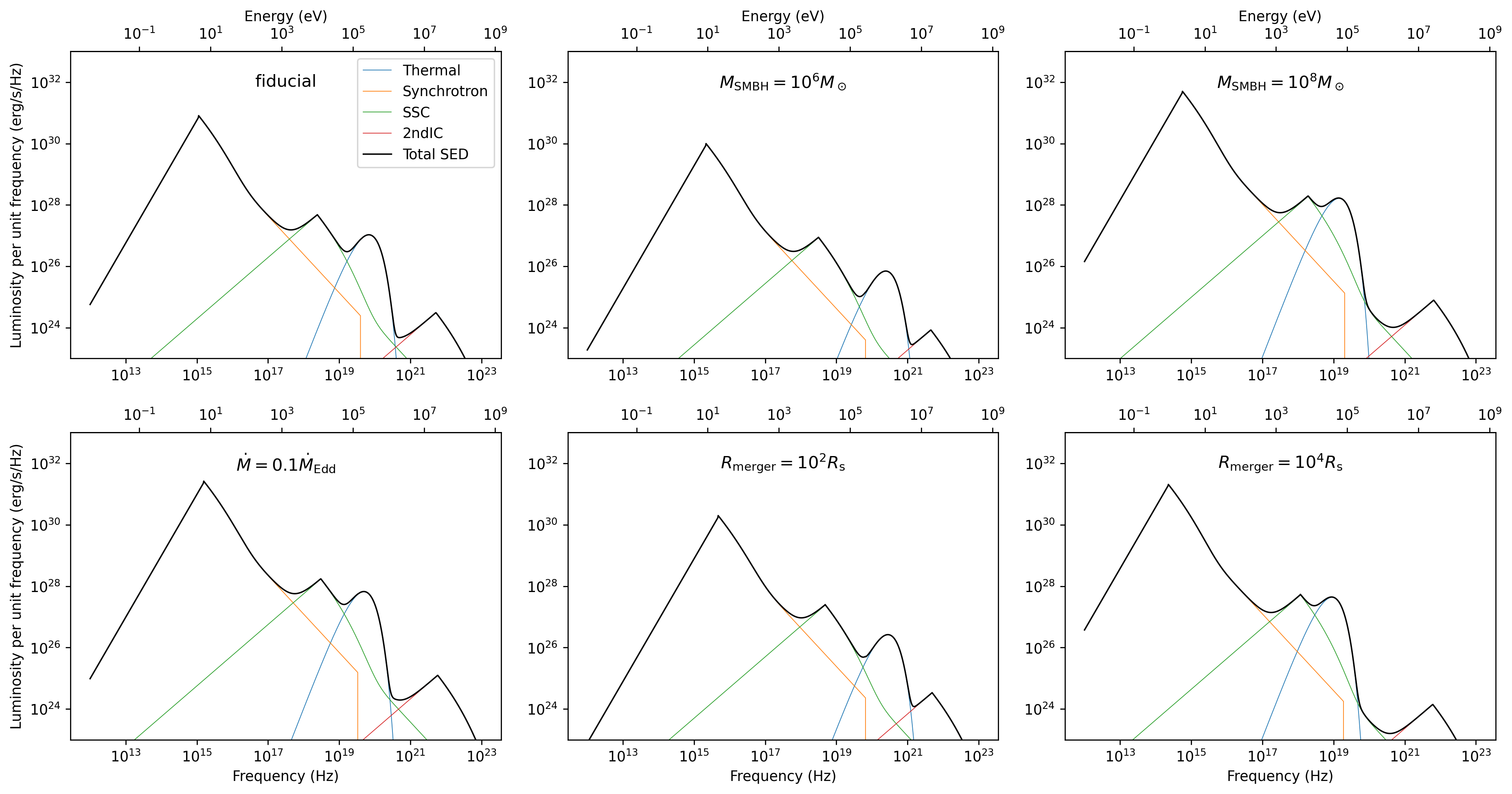} 
\caption{The expected SED of a GRB originating from a BBH merger in AGN disks. In different panels, only the parameter shown in the corresponding panel is altered, while the remaining parameters remain consistent with the fiducial model. In this study, the fiducial parameters consist of: $M_{\rm SMBH} = 10^7 M_\odot$, $\dot{M}=0.01\dot{M}_{\rm Edd}$, $\tilde{\alpha} = 0.05$, $\tilde{R}=10^3 R_{\rm s}$, $m_{\rm BH} = 150 M_\odot$.
}
\label{fig:prompt-absorption}
\end{figure*}
The observed GRB-like event following LVK S241125n shows a distinctive feature in its prompt emission, with a photon index of -2.2 in the 15-350 keV range. This index differs from the typical photon index of around -1 to -1.5 seen in short GRBs. We demonstrate that these features closely align with the scenario of a BBH merger in an AGN disk.

To simulate the SED of a GRB explosion in an AGN accretion disk, we consider the newly formed BH accreting material from the AGN disk, which generates a jet. This jet interacts with the AGN disk material and ultimately erupts from the disk surface, as detected by the instruments. Our calculations are based on the framework established by \citet{2023ApJ...950...13T}.

We apply the standard disk model \citep{1973A&A....24..337S}, as summarized in \citet{kato1998black}, for the AGN accretion disk. This model is similar to that described in \citet{2003MNRAS.341..501S} (SG), with the key difference being their use of tabulated opacities. The fiducial parameters chosen for our computations include an accretion rate of $\dot{M} = 0.01 \, \dot{M}_{\text{Edd}}$ and a viscosity parameter of $\tilde{\alpha} = 0.05$, where $\dot{M}_{\text{Edd}}$ represents the Eddington accretion rate; the GRB explosion occurs at a distance of $10^3 R_{\rm s}$, where $R_s$ is the Schwarzshild radius of the center supermassive balck hole (SMBH). The outer boundary of the AGN disk is set at $2\times10^4 R_{\rm s}$, considering that the outer region might be gravitationally unstable.
The accretion rate of the newly merged BH is given by the Bondi–Hoyle–Lyttleton (BHL) accretion rate, modified by the disk's scale height and tidal forces, combined with an enhancement factor $f_{\text{acc}} = 15$ due to the kick resulting from the merger. To calculate the jet power, we assume that the jet luminosity is proportional to the accretion rate and a certain conversion efficiency of $\eta_{\rm i}=0.5$. The interaction of the jet with the AGN disk material is considered, and the velocity of the jet head is calculated using the formulas from \citet{2022ApJ...927...41T}. The radiation is computed following \citet{2023ApJ...950...13T}, where the jet, emerging from the disk surface, can simultaneously produce both thermal and non-thermal components. The latter includes synchrotron radiation, synchrotron self-Compton (SSC), and second-order inverse Compton scattering (2ndIC). The plasma parameters related to non-thermal emission are set to be the same as those in \citet{2023ApJ...950...13T}. 

Meanwhile, AGN disks are promising environments for intermediate mass black hole (IMBH) mergers and growth due to their unique dynamical conditions, high escape velocities, and high gas densities \citep[e.g.,][]{2025arXiv250813412D}. Thus, for the mass of the merging BHs, we use the merger product mass from GW190521 as the reference model, specifically \( 150 M_{\odot} \). To ensure its validity for S241125n, we use the luminosity distance to estimate the lower limit of the detectable mass through the scaling law. The inspiral signal-to-noise ratio (SNR) scales as 
\begin{equation}
\text{SNR} \propto \frac{M_{\text{c}}^{5/6}} {d_{\rm L}},
\end{equation}
where \( M_{\text{c}} \) is the chirp mass \citep{cutler1994gravitational}. For example, the chirp mass of an equal-mass binary neutron star is given by $M_{\text{c}} = m_{\rm NS}/2^\frac{1}{5} \sim 1.2M_\odot$. For this typical NS merger, the maximum range achievable by LIGO is 177 Mpc \citep{2024arXiv241114607C}. Therefore, for a source at \( d_{\rm L} = 4173 \) Mpc, the required chirp mass scales accordingly, leading to an estimated chirp mass of approximately $M_c \approx 1.2M_\odot \times \left(\frac{4173}{177}\right)^\frac{6}{5} \approx 53 M_\odot$. Converting to total mass for an equal-mass binary, this results in a required total mass of approximately $m_{\rm BH} = 2\times M_c \times 2^\frac{1}{5} \approx 122M_\odot$. Therefore, using \( 150 M_{\odot} \) as the reference mass is reasonable.

Fig. \ref{fig:prompt-absorption} shows the expected SED of the prompt emission from a GRB originating in an accreting BH remnant in AGN disks. We consider different parameter settings relative to the fiducial case, including various masses of SMBHs, the SMBH’s accretion rate, and different locations of the GRB explosion in the AGN disk. Although the parameters vary widely, the outcomes are qualitatively similar and share common characteristics. Specifically, in the low-energy band, synchrotron radiation dominates, while around 10 keV, the contribution is mainly from SSC, and in the gamma-ray range, it is primarily from 2ndIC. In the energy range of $\sim10-100\ \rm keV$, which corresponds to the observational band of Swift-BAT, thermal radiation dominates, and its high-energy side decreases rapidly with frequency, resulting in a steep spectral index. The thermal peak may also partially overlap with the SSC contribution and serve as the main energy radiation band. 

Having described the mechanism and general observational features of GRBs potentially produced by BBH mergers in AGN disks, we now turn our attention back to S241125n.
Generally, the break power law with a photon index of -2.2 in the observational energy range (15-350 keV in the observer frame) can be interpreted as the high-energy wing of thermal radiation or as the high-energy wing of SSC.
We also consider the SG disk model and the disk model proposed by \citet{thompson2005radiation} (TQM). A very similar overall SED and photon index feature in the 15-300 keV range is also found using the SG disk model. However, generally lower luminosities are found using the TQM disk model, which is not consistent with S241125n.

The time delay between GW and GRB, and the isotropic equivalent breakout luminosity of the relativistic shock are given by \citep{2023ApJ...950...13T}:
\begin{equation}
t_{\rm delay} \sim (1+z) \frac{\tilde{H}}{4\gamma_{\rm sf}^2 c} = 11.264\ {\rm s},
\end{equation}
and 
\begin{equation}
\begin{aligned}
L_{\rm breakout} &\sim \pi f_{\rm beaming} \theta_{\rm j}^2 \tilde{H}^2 \tilde{\rho} v_{\rm FS} \gamma_{\rm FS}^2 c^2 \frac{\gamma_{sf,f}}{4\gamma_{\rm sf}} \\ &= 1.01\times 10^{51} \rm erg\ s^{-1},
\end{aligned} 
\end{equation}
where $f_{\rm beaming}=2\gamma_{\rm sf,f}^2$; $\gamma_{\rm sf}$ and $\gamma_{\rm sf,f}$ are the Lorentz factor and the final Lorentz factor of the shocked fluid, respectively; $v_{\rm FS}$ and $\gamma_{\rm FS}$ are the velocity and Lorentz factor of the shock.  By adjusting the final Lorentz factor of the shocked fluid to \( \gamma_{\text{sf},f} = 15 \) (corresponding to a final Lorentz factor of the shock of $\gamma_{\rm FS,f} \approx 40$), we can estimate the thickness and density of the AGN disk at the explosion site as $\tilde{H} \approx 5.69\times 10^{12} \rm cm$ and $\tilde{\rho} \approx 4.20\times 10^{-9} \rm g\ cm^{-3}$, assuming an opening angle of $\theta_{\rm j} = 0.1$. The burst duration predicted by the model is then estimated as \(\Delta t = (1+z)\tilde{H}/(2 \gamma_{\text{sf},f}^2 c) \approx 0.729\,\mathrm{s}\), which falls within the range of short GRBs. These inferred thickness and density values are realized for a standard AGN disk with a common accretion rate of $\dot{M} = 0.01 \, \dot{M}_{\text{Edd}}$, a central mass of $M_{\rm SMBH} = 10^7M_\odot$, and $\tilde{\alpha} = 0.05$ at a radius of $R_{\rm merger} \sim 8\times10^2 R_{\rm S}$. 

The AGN disk material not only affects the prompt emission of the GRB but also influences the afterglow. The location of the afterglow may be displaced from the mid-plane of the disk, resulting in a smaller column density of disturbing gas.

\subsection{Afterglow}
\label{sec:3.2}
\subsubsection{X-ray: Photoelectric Absorption}

\begin{figure}
\centering
\includegraphics[width=1\linewidth]{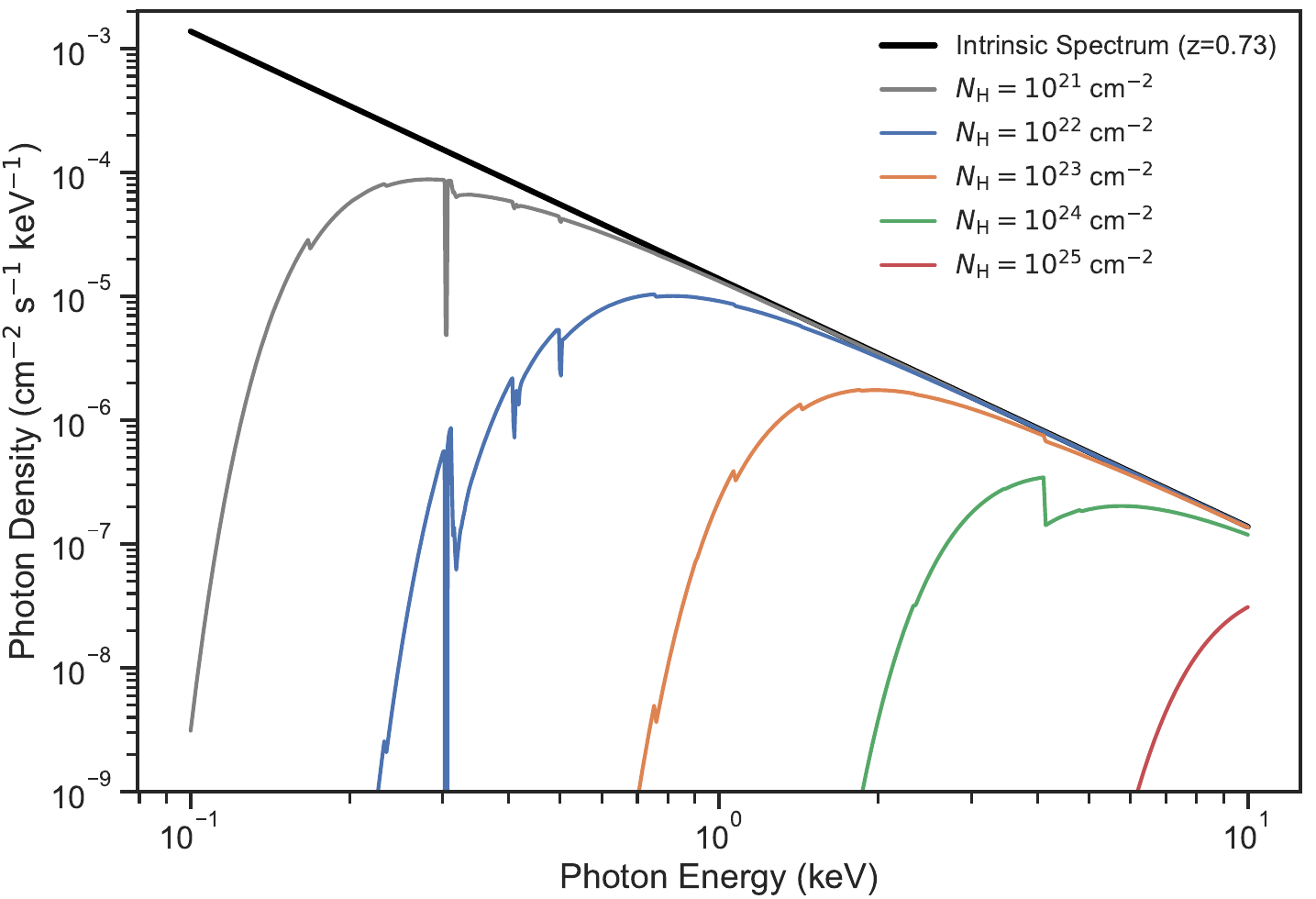} 
\caption{X-ray absorption effects on a power-law spectrum of initial photon index = 2 emitted from redshift $z = 0.73$ and observed at $z=0$.  The black line shows the intrinsic spectrum with a total flux of $2.20 \times 10^{-13}$ erg cm$^{-2}$ s$^{-1}$ from 0.1 keV to 10 keV. Colored lines show the absorbed spectra for different neutral hydrogen column densities ($N_\mathrm{H}$) from $10^{21}$ to $10^{25}$ cm$^{-2}$. As $N_\mathrm{H}$ increases, the absorption cutoff energy shifts toward higher energies and the observed spectrum becomes harder, resulting in a flatter spectral shape above the cutoff. At $N_\mathrm{H} = 10^{23} \, \mathrm{cm}^{-2}$, 75.5\% of the intrinsic flux is transmitted ($1.66 \times 10^{-13} \, \mathrm{erg} \, \mathrm{cm}^{-2} \, \mathrm{s}^{-1}$), while at $N_\mathrm{H} = 10^{24} \, \mathrm{cm}^{-2}$, only 40.0\% is transmitted ($8.81 \times 10^{-14} \, \mathrm{erg} \, \mathrm{cm}^{-2} \, \mathrm{s}^{-1}$)}
\label{fig:xray-absorption}
\end{figure}

X-ray photons are absorbed when their energy exceeds the ionization energy of bound electrons in atoms or ions. The dominant process is photoelectric absorption, where a photon interacts with an atom, ionizing it and ejecting an electron. The optical depth $\tau$ for photon energy $\epsilon$ is usually approximated by
\begin{equation}
\tau(\epsilon) = N_{\rm H} \cdot \sigma_\mathrm{eff}(\epsilon)
\end{equation}
where $N_{\rm H}$ is the hydrogen column density, and $\sigma_{\rm eff}$ is the effective photoelectric cross-section, including contributions from hydrogen, which dominates at low energies (below $\sim 0.3$ keV), helium, which is important for soft X-rays ($\sim 0.3-1$ keV), and heavier metals, which contribute at higher energies due to K-shell and L-shell edges \citep{1983ApJ...270..119M, 1992ApJ...400..699B}. 

From the observation, we may find the absorption suppressing the soft X-ray flux ($\epsilon < 1$ keV) for high column densities ($N_H > 10^{21}\,\mathrm{cm}^{-2}$) \citep{2010MNRAS.402.2429C, 2010MNRAS.401.2773S}. At higher energies, the spectrum is less affected due to the rapid decline of the photoelectric cross section ($\sigma_\mathrm{eff} \propto \epsilon^{-3}$). We may also find sharp drops in flux at specific energies corresponding to the K-shell and L-shell binding energies of heavy elements (e.g., the oxygen K-edge at 0.54 keV, silicon K-edge at ~1.8 keV, and iron K-edge at ~7.1 keV) \citep{2000ApJ...542..914W}.

AGN disks are generally dense, of $N_H \sim 10^{22} - 10^{25} \, \mathrm{cm}^{-2}$ \citep{1999ApJ...522..157R}, the exact column density depends on the geometry, density profile, and distance from the BH \citep{1993ARA&A..31..473A}. The inner parts of the disk (few parsecs or closer) are dominated by high-density gas ($n \sim 10^8 - 10^{12} \, \mathrm{cm}^{-3}$), leading to column densities as high as $N_H > 10^{24} \, \mathrm{cm}^{-2}$. The outer regions of the disk, extending to kiloparsec scales, are less dense, with $N_H \sim 10^{21} - 10^{23} \, \mathrm{cm}^{-2}$, this region may overlap with the torus or host galaxy, where additional gas and dust contribute to absorption. The circumnuclear torus surrounding the AGN can have column densities of $N_H \sim 10^{23} - 10^{25} \, \mathrm{cm}^{-2}$, depending on the inclination angle and torus geometry \citep{1999agnf.book.....K,2008NewAR..52..274E,2015ARA&A..53..365N}.

The high $N_H$ of the AGN disk will significantly absorb the GRB's afterglow emissions, especially in the soft X-ray band ($< 10 $ keV), the observed spectrum can be hardened with a low-energy cutoff \citep{2010MNRAS.402.2429C, 2010MNRAS.401.2773S}. To simulate, we adopt the ISM grain absorption models from XSPEC\footnote{\url{https://heasarc.gsfc.nasa.gov/xanadu/xspec/manual/node275.html}}, of which the  elements abundances are from \citet{2000ApJ...542..914W} and the cross sections come from \citet{1996ApJ...465..487V}. We assume the initial spectrum follows a power-law\footnote{\url{https://heasarc.gsfc.nasa.gov/xanadu/xspec/manual/node220.html}} with a typical photon index $2$ and a normalization $0.003$ cm$^{-2}$ s$^{-1}$ keV$^{-1}$ at $1$ keV emitted at $z=0.73$, and adopt the solar metallicity. The absorbed X-ray spectra for column densities spanning $10^{21} - 10^{25} \, \text{cm}^{-2}$ observed at $z = 0$ are presented in Figure \ref{fig:xray-absorption}. The results reveal a systematic shift in the low-energy cutoff with increasing $N_H$, rising from $\sim 0.2 \, \text{keV}$ at $N_H = 10^{21} \, \text{cm}^{-2}$ to over 10 keV at $N_H = 10^{25} \, \text{cm}^{-2}$. Concurrently, the spectral profile flattens, with the power-law index $\sim 0$ at \(N_H = 10^{24} \, \text{cm}^{-2}\). These results demonstrate that significant hardening of the apparent photon index can occur if the afterglow is significantly absorbed. The photon index of the afterglow observed by EP (\(0.43^{+0.76}_{-0.74}\)), while substantially harder than the canonical value of 2, is consistent with scenarios involving absorption by an AGN disk of $N_H \simeq 10^{23} - 10^{24} \, \text{cm}^{-2}$. 

\subsubsection{Infrared, Optical and UV: Extinction}

\begin{figure*}
\centering
\includegraphics[width=0.9\linewidth]{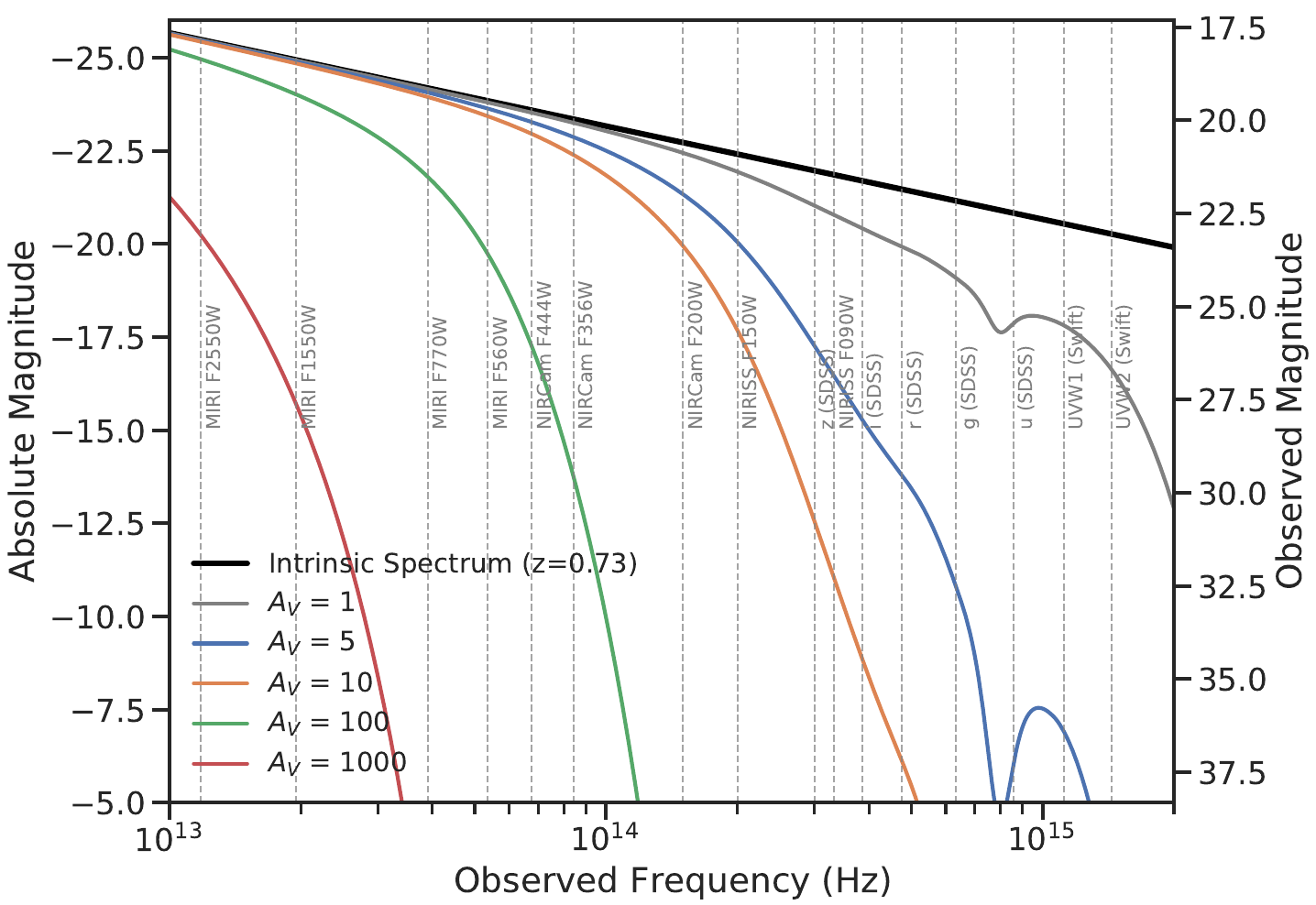} 
\caption{Extinction effects of infrared, optical and UV bands from a synchrotron spectrum emitted at z = 0.73 and observed at z=0. The black line shows the intrinsic power-law spectrum of spectral index = 1 (photon index = 2) for typical afterglow of short GRB at $\sim 1$ day, while colored lines represent the same spectrum affected by different levels of extinction ($A_V = 1-1000$). Vertical dashed lines indicate the central frequencies of various astronomical filters from SDSS optical bands (u, g, r, i, z), JWST near- and mid-infrared instruments (NIRISS, NIRCam, MIRI) and SWift-UVOT UV bands. The extinction becomes increasingly severe at higher frequencies (shorter wavelengths) and higher $A_V$ values, assuming the wavelength-dependent nature of dust extinction following the Fitzpatrick-Massa (2007) extinction law. This demonstrates that while UV and optical emission from a GRB afterglow may be heavily attenuated by AGN dust, the signal can remain observable in the infrared.}
\label{fig:optical-extinction}
\end{figure*}

Figure \ref{fig:optical-limits} presents the optical observations of S241125n, which consist only of upper limits from various telescopes, as well as the comparison with other short bursts. No definitive optical counterpart from S241125n was identified, implying that the GRB’s optical afterglow was either intrinsically faint or heavily obscured. 

One of the primary reasons for the absence of optical afterglows is strong dust extinction. In particular, GRBs occurring in high-density environments, such as star-forming regions or AGN disks, may be completely obscured in the optical band. From previous observations, a fraction of GRBs have no detected optical afterglows, leading to their classification as ``dark GRBs'' \citep{2004ApJ...617L..21J}, many of which have been confirmed to be associated with host galaxies that exhibit high  $N_H$ \citep{2010AIPC.1279..144G,2009AJ....138.1690P}.

The total extinction is characterized by the visual extinction parameter $A_V$, which is related to the hydrogen column density  $N_H$. For typical Milky Way-like gas-to-dust ratios, the relation is  approximated as \citep{1978ApJ...224..132B,1995A&A...293..889P,2012AAS...22052301F}
\begin{equation}
A_V \text{(Milky Way)} \approx 5 \times 10^{-22} N_H
\end{equation}
Observations of AGN suggest that the dust-to-gas ratio is often lower than in the Milky Way due to the destruction of dust grains by strong radiation fields, shocks, and turbulent environments in the AGN vicinity, expressed as \citep{2001A&A...365...37M, 2016A&A...586A..28B}
\begin{equation}
A_V \text{(AGN)} \approx 5 \times 10^{-23} N_H.
\end{equation}
For example, an AGN disk with $N_H \sim 10^{23}-10^{24} ~\text{cm}^{-2}$, we have
\begin{equation}
A_V \approx 5 - 50
\end{equation}

Using the \citet{2007ApJ...663..320F} extinction law, the estimated impact on the GRB afterglow is shown in Figure \ref{fig:optical-extinction}. The extinction is more severe at shorter wavelengths, with UV and optical photons being heavily absorbed. At $A_V > 5$, the flux in the UV band is completely suppressed, while in the optical bands, the attenuation exceeds several magnitudes, making the GRB afterglow undetectable in deep optical surveys. Even though powerful jets in AGN disks may lead to delayed UV/optical flares \citep{2025arXiv250516390C}, extinction likely suppresses these signals, which is consistent with the observational constraints from WFST. However, the infrared flux remains less affected, particularly in mid-infrared wavelengths. 

Infrared observations are crucial in such cases. JWST’s NIRCam and MIRI instruments \citep{2006SSRv..123..485G}, operating at wavelengths from 1 to 25 µm, are capable of detecting GRB afterglows even in highly extincted environments. From Figure \ref{fig:optical-extinction}, we estimate that at $A_V \sim 10$, the afterglow would remain observable in JWST’s mid-infrared bands (e.g., MIRI F560W and F770W). Assuming the infrared afterglow follows a power-law decay with an index $-1.5$ \citep{2018ApJS..234...26L, 2025ApJ...978...51D}, and starts at $\sim 20$ mag at 1 day, it is expected to fade to $\sim 25.5$ mag at 1 month and $\sim 27$ mag at 3 months. JWST’s MIRI instrument, capable of detecting sources down to $28-29$ mag in deep exposures, should still be able to observe the afterglow within the first months. However, this event lacks timely follow-up observation; similar events deserve follow-up observations in future. 

In summary, the lack of an optical detection for S241125n is consistent with strong extinction expected in a dusty environment, which is aligned with the inference from the X-ray analysis. 

\section{Discussion on host galaxy}
\label{sec:4}
\begin{figure*}
\centering
\includegraphics[width=0.9\linewidth]{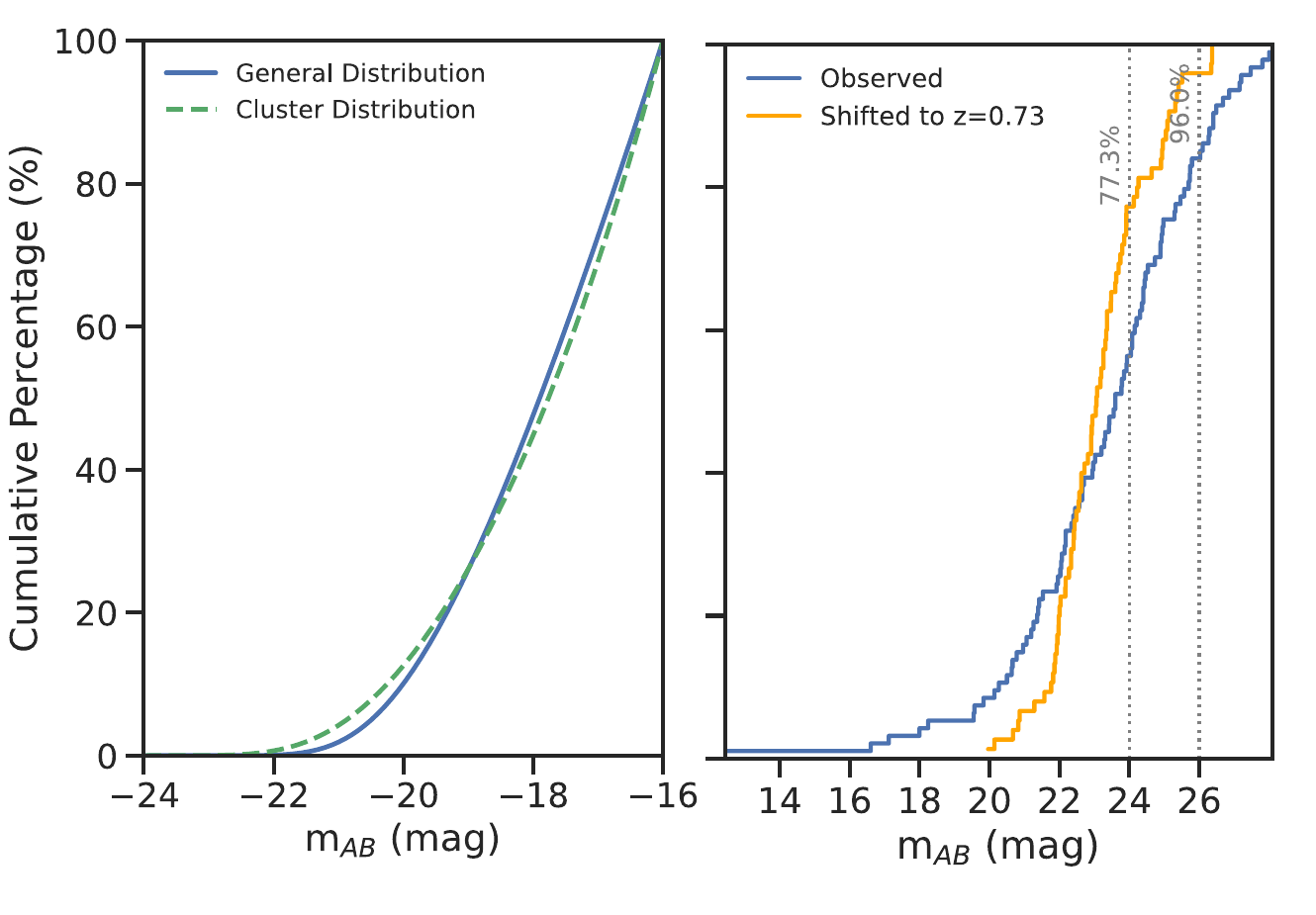} 
\caption{\textbf{Left:}  Cumulative percentage from integrating the luminosity function for the general galaxy distribution (solid blue line) and the composite cluster galaxy distribution (dashed green line). \textbf{Right:}  Cumulative distributions for the host galaxies of short GRBs. The observed AB magnitudes (blue step line, sample size = 94) and the AB magnitudes shifted to the redshift of 0.73 (orange step line, sample size = 75). Vertical grey dotted lines mark $m_{AB}$ = 24 and 26, the corresponding cumulative percentages are annotated, $77.3\%$ and $96.0\%$, respectively.}
\label{fig:ratio}
\end{figure*}

Once a GRB is localized with high accuracy, observers use deep imaging to look for galaxies in this region \citep{2002AJ....123.1111B}. The likely host is the galaxy that is closest to the afterglow position. Spectroscopy provides the galaxy's redshift, which is then compared to the afterglow's redshift, a match confirms the association. For S241125n, the sky region corresponding to the 5-arcminute uncertainty provided by BAT is relatively large, and no publicly available data has conducted a deep search for the host galaxy in this region following the GRB event.

Comparing catalogs from surveys is another method to identify a host galaxy. These surveys offer photometric data and sometimes spectroscopic data over large areas of the sky, which helps in cross-matching the GRB position with known galaxies. Unfortunately, no host galaxy was found within the uncertainty radius of S241125n's localization in these catalogs. This raises a question: Did the surveys fail to detect S241125n’s host galaxy due to instrumental limitations, or is S241125n truly extragalactic?


For the optical surveys, the limiting magnitude of SDSS typically reaches $\sim 22$ mag \citep{2000AJ....120.1579Y}, Pan-STARRS can detect sources down to $\sim 23$ mag \citep{2016arXiv161205560C}, while DES and DESI reaches about $\sim 24$ mag \citep{2021ApJS..255...20A, 2016arXiv161100036D}. In the near-infrared, 2MASS and WISE have limits of approximately $\sim 16.5$ mag \citep{2006AJ....131.1163S, 2010AJ....140.1868W}, the spectroscopic survey of GAMA can reach $\sim 20$ mag \citep{2015MNRAS.451.3249A}. Therefore, if the galaxy's observed magnitude is fainter than 24 mag, it would exceed the observational limits of these surveys and would not be found in the corresponding catalogs. For some deep field surveys, such as Hyper Suprime-Cam Subaru Strategic Program (HSC-SSP) \citep{2018PASJ...70S...8A}, Hubble Ultra Deep Field (HUDF) \citep{2006AJ....132.1729B}, UltraVISTA \citep{2012A&A...544A.156M}, and Spitzer Extended Deep Survey (SEDS) \citep{2013ApJ...769...80A}, the observational limiting magnitudes exceed 24 mag, with some reaching beyond 26 mag. However, due to their limited sky coverage, these surveys do not encompass the location of S241125n.

Let us first consider intrinsic faintness. For a galaxy with absolute magnitude $M$ in a given band, the expected apparent magnitude is computed via the distance modulus $m = M + \mu + K$, where $\mu = 5 \log_{10}(d_L/10~\mathrm{pc})$ and $K$ is the $k$-correction. For $z=0.73$ the luminosity distance $d_L$ is 4173 Mpc, and the corresponding $m = 43.2$. For a typical galaxy with $\sim -21$ and a sub-luminous galaxy $\sim -19$, the expected apparent magnitude (ignoring $K$-corrections for simplicity) is $22.2$ mag and $24.2$ mag, respectively.

Next is to consider dust extinction. Typical optical extinction $A_V$ is about $0.5$ to $1.0$ mag. In galaxies with strong star formation or in starburst galaxies, $A_V$ can be $1$ – $2$ mag or more \citep{1994ApJ...429..582C, 2000ApJ...533..682C}. In AGNs, extinction varies widely. For obscured (Type 2) AGNs, $A_V$ can be much higher, frequently 5 mag and sometimes reaching 10 mag or above \citep{2003AJ....126.1131R, 2024A&A...687A.159B}. If we take $A_V \sim 1$, the apparent magnitude would be dimmed to $23.2$ mag and $25.2 $mag. Furthermore, cosmological surface brightness dimming scales as $(1+z)^4$ \citep{1930PNAS...16..511T}. At $z=0.73$ this factor is $\sim 9$. Extended, low-surface-brightness features of the host might then fall below the detection threshold even if the integrated magnitude is bright enough. 

Using the Schechter luminosity function \citep{1976ApJ...203..297S, 2001AJ....121.2358B}, one can estimate the fraction of galaxies with luminosities below a given threshold. 
Figure \ref{fig:ratio} demonstrates the cumulative percentage obtained by integrating the luminosity function for both the general galaxy distribution and for a composite cluster galaxy distribution. It shows over $70\%$ of galaxies are fainter than $-19$ mag and over $90\%$ are fainter than $-20$ mag. Therefore, in a survey limited to about 24 mag, the majority of galaxies at $z=0.73$ would naturally be missed.

From the statistics of GRB observations, \citet{2017MNRAS.467.1795L} conducted a Hubble Space Telescope (HST) WFC3/F160W Snapshot survey of $39$ long-duration GRB host galaxies and concluded that the hosts are statistically fainter compared to typical field galaxies. $17$ hosts have magnitudes fainter than $24$ mag, representing $43.6\%$ of the total sample. Additionally, $5$ hosts were not detected, suggesting they may be even fainter than $26$ mag or obscured.

\citet{2022ApJ...940...56F} analyzed the host galaxies of 94 short GRBs, with the cumulative percentage of different observed magnitudes shown as the blue line in the right panel of Figure \ref{fig:ratio}. We then shifted these host galaxies to  $z = 0.73$  to assess their observability. Among these 94 short GRBs, 19 lack redshift measurements due to their faintness, with a lower magnitude limit ranging from $23.6$ to $28.1$. For the remaining 75 bursts, the results after redshift transformation are depicted as the orange line in the right panel of Figure \ref{fig:ratio}. It shows that, after shifting to  $z = 0.73$, $17$ host galaxies exceed 24 mag. Considering the $19$ faint host galaxies with undetected redshifts, this suggests that $18\% - 38\%$ of the host galaxies are fainter than 24 mag at $z = 0.73$. This observed rate is lower than the value calculated from the Schechter luminosity function, which is likely due to the fact that higher-luminosity galaxies contain more stars and thus have a higher probability of producing GRBs. 

S241125n, as a special GRB, whether its host galaxy follows the typical trends of ordinary GRBs remains uncertain. Nevertheless, based on the observations and calculations above, we can reasonably infer that the non-detection of the host galaxy of S241125n in surveys is expected, with a probability ranging from a minimum of $\sim 20\%$ to more than $50\%$. 

Although we have considered general cases, the detectability of host galaxies is also related to BH formation processes if the GRB emissions originate from AGNs. This is because the masses of the galaxies are biased (being less than $10^7 M_\odot$ or greater than $10^7 M_\odot$) depending on the BH formation scenarios. If these BHs are captured from nuclear star clusters, theory predicts that the SMBH mass would likely be lower, largely because BHs in a nuclear star cluster can be more easily captured by an AGN disk around a lower-mass SMBH \citep{2025arXiv250419570X}. Conversely, if they are formed by in-situ star formation, merger rates roughly follow the AGN luminosity function, with most mergers occurring in SMBHs around $10^7$–$10^8 M_\odot$. 
Thus, some caution is required in analyzing the detectability, and 
the latter scenario could be constrained due to the lack of detected host galaxies.

Although the null hypothesis that GW (LIGO) and EM (Swift or/and EP) are completely unrelated is disfavored, we cannot completely rule out that they are unrelated since the redshift of the GRB has not been determined by its spectrum or its host galaxy. In the case that they are unrelated, the GRB might originate from a greater distance. Additionally, GW has poor localization, so it could still be at a greater distance even if they are associated. These scenarios could make the detection of a host galaxy even less likely. 

\section{Conclusions}
\label{sec:5}
In this work, we explored GRBs emerging from binary BH mergers in AGN disks, focusing on both the features of the prompt emission and the absorption and extinction in the afterglow. LVK S241125n is a known rare GW candidate from a binary BH merger detected with a putative coincident short GRB in multiwavelength observations, with a joint GW+BAT+EP $\mathrm{FAR}_{\rm triple}$ of 1/30 yr. This GRB-like event exhibits specific spectral index in both the prompt and afterglow phases, and thus serves as an interesting test model of a binary BH merger event in an AGN disk.

The merger product of a binary BH in an AGN disk can significantly exceed the Eddington limit, leading to the formation of relativistic jets. These jets interact with the disk material, creating shock waves that eventually break out at the disk surface. The breakout produces corresponding thermal and non-thermal radiation. The photon index of -2.2 in the prompt phase of the GRB following LVK S241125n may be interpreted as the high-energy wing of the thermal component or as the SSC contribution from the non-thermal component. In contrast, the unusually hard photon index in the afterglow phase can be attributed to absorption by the high-density environment. Furthermore, the time delay between the GW and GRB signals, as well as their luminosities, can be fitted within reasonable parameter ranges consistent with a binary BH merger event in an AGN disk. However, the non-detection of an optical counterpart suggests strong extinction, complicating the identification of the host galaxy and highlighting the need for more sensitive infrared observations.

Future studies of S241125n and similar events could provide deeper insights into the fundamental physics of BH mergers and their role in the broader cosmic landscape, potentially uncovering new connections between GWs, electromagnetic signals, and the host environments of these extraordinary phenomena.

\section*{Acknowledgment}
We sincerely thank the referee for her/his interest in our work and for the constructive suggestions, which have helped make the article more complete and solid. We Acknowledge the discussion with Prof. Antonio Enea Romano on the potential alternative methods for model verification following the public release of future gravitational wave data. 
We thank Dr. Liang-Gui Zhu for the discussion regarding the gravitational-wave skymap plot.
YFY is supported by National Natural Science Foundation of China (Grant No. 12433008, 12393812),
National SKA Program of China (Grant No. 2020SKA0120300) and
the Strategic Priority Research Program of the Chinese Academy of Sciences (Grant No. XDB0550200, XDB0550300).

\appendix

\section{Simulation Test for S241125n Duration}

We conducted a forward-folding simulation to evaluate if a GRB with peak isotropic luminosity $L_{\mathrm{p}} = 3 \times 10^{50}\,\mathrm{erg\,s^{-1}}$ at a redshift of $z = 0.73$ and intrinsic rest-frame pulse width $T_{\mathrm{RF}} = 2\,\mathrm{s}$ would indeed appear as an event of observed $T_{90} \sim 0.5\,\mathrm{s}$ recorded by the Swift--BAT. The simulation follows a standard procedure for Swift-BAT\footnote{\url{https://swift.gsfc.nasa.gov/analysis/threads/batsimspectrumthread.html}}.

\begin{figure*}
\centering
\includegraphics[width=0.9\linewidth]{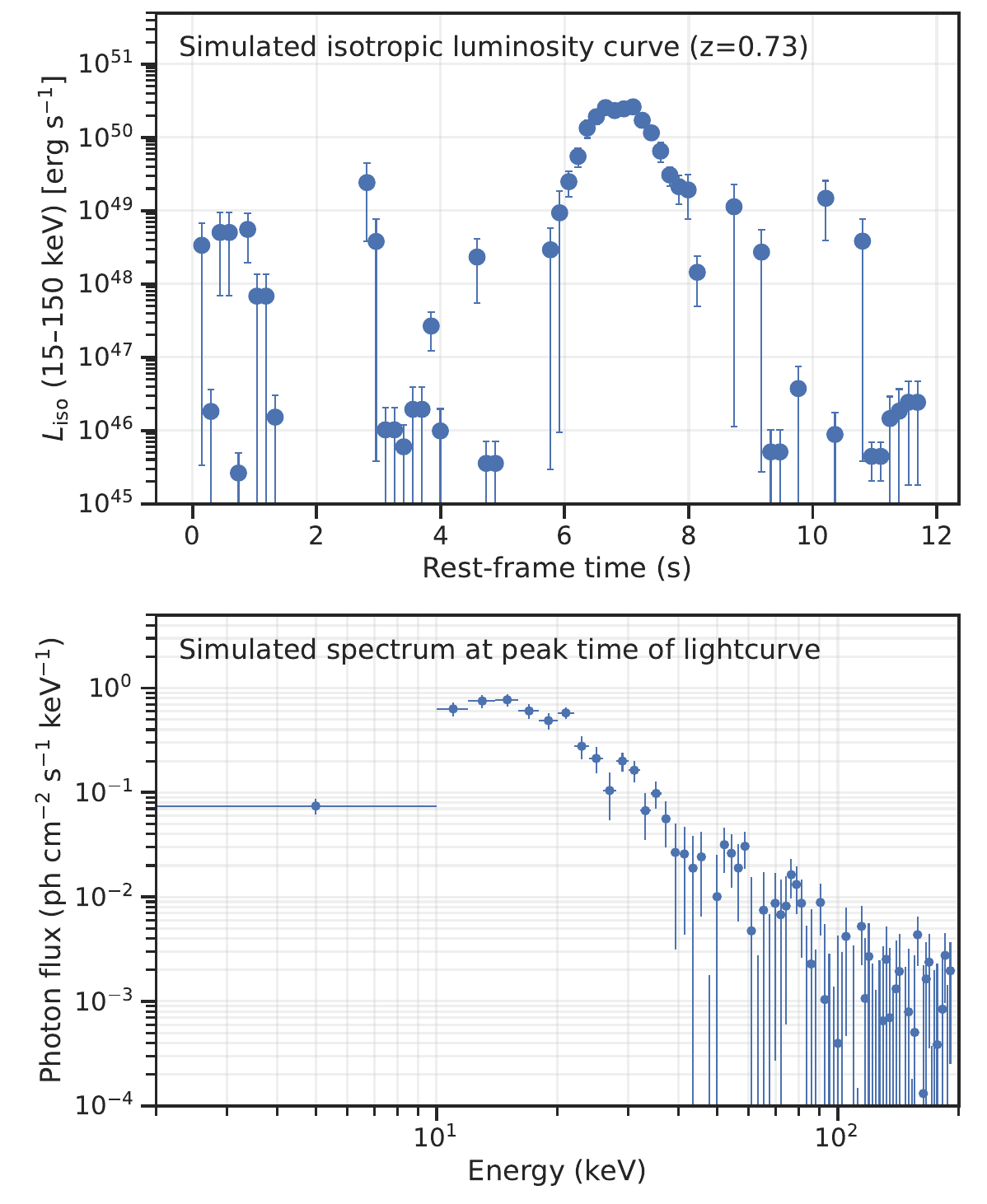} 
\caption{Simulation results for the Swift–BAT detection capability for a $2$~s duration Norris profile pulse of peak luminosity $\sim 3 \times 10^{50}~\text{erg s}^{-1}$  at redshift $z=0.73$. \textbf{Upper}: The observed isotropic luminosity curve in the rest frame (15–150 keV). \textbf{Lower}: Simulated photon flux spectrum at the peak luminosity time, illustrating the spectral energy distribution folded through the BAT response.}
\label{fig:bat-simulation}
\end{figure*}

Due to the absence of original Swift-BAT event data for S241125n, background estimation relied on standard models from the HEASARC Calibration Database (CALDB)\footnote{\url{https://heasarc.gsfc.nasa.gov/docs/heasarc/caldb/swift/}}. The CALDB background model for BAT comprises long-term observations of blank-sky fields, integrating instrumental background and cosmic diffuse X-ray emission averaged over time. These standardized models reside in the \texttt{bcf/bkg} branch of the CALDB directory and can be utilized by tools like \texttt{batphasimerr} to estimate Poisson errors for simulated spectra. The use of CALDB backgrounds instead of real observational data typically introduces uncertainties of approximately $\pm 30$--$50\%$ in counts, acceptable for approximate detectability studies, particularly for long duration events \citep{markwardt2007swift, 2013ApJS..209...14K}.

For the spectral model, we adopted a cutoff power-law spectrum to be consistent with the observation, with a fixed photon index $\Gamma = 2.2$ and a peak energy $E_{\mathrm{peak}} = 49\,\mathrm{keV}$. The normalization  of the spectrum follows a Norris pulse profile \citep{1996ApJ...459..393N} defined as
\begin{equation}
A(t) = A_0 \exp\left(-\frac{(t - t_0)^2}{2w^2}\right),
\end{equation}
where $A_0$ is the amplitude, $t_0$ is the pulse peak time, and $w$ is the pulse width. For this simulation, we set $t_0 = 12\,\mathrm{s}$ (observer frame) and pulse width $w = 1\,\mathrm{s}$, resulting in an observed $T_{90}$ of approximately $3.5\,\mathrm{s}$, which corresponds to a rest-frame duration of about $2\,\mathrm{s}$. The simulated light curve consists of 80 time bins, each spanning $0.256\,\mathrm{s}$. The simulation steps involved
\begin{enumerate}
\item Generating photon spectra using XSPEC.
\item Creating a background-free PHA file with the XSPEC command \texttt{fakeit}.
\item Folding spectra through the on-axis BAT response matrix file (RMF) \texttt{swbresponse20030101v007.rsp}.
\item Applying \texttt{batphasimerr} with parameters \texttt{bkgfile=CALDB} to incorporate BAT background and Poisson noise.
\end{enumerate}

This approach produces per-bin count rates and statistical errors. The energy flux $F_i$ in the 15--150 keV band is then calculated from the spectral model via \texttt{AllModels.calcFlux}. Luminosities $L_i$ are derived from the flux, $L_i = 4\pi D_{\!L}^{2}\,F_i$,
where the luminosity distance $D_{\!L} = 4173$ Mpc.

If S241125n were indeed a disguised long-duration GRB with a tail dropping below detectability thresholds, our simulation would predict a significantly shorter observed duration compared to its intrinsic duration of $2\,\mathrm{s}$. However, our simulated light curve shown in Fig. \ref{fig:bat-simulation} demonstrates that nearly all time bins remain above the background detection threshold of $\sim 10^{49}~\text{erg s}^{-1}$. Consequently, the short observed duration of approximately $0.5\,\mathrm{s}$ cannot be attributed to Swift-BAT's limited sensitivity and is therefore intrinsic to the GRB itself.

Swift–BAT employs trigger algorithms based on accumulated counts exceeding the background. The total photon count of short GRBs, integrated over their brief duration, is typically smaller compared to that of longer-duration GRBs. Consequently, short GRBs have inherently lower signal-to-noise ratios (SNR). Therefore, short bursts may fail to trigger the standard Swift–BAT algorithms, even when a longer burst with identical peak flux would be successfully triggered. However, untriggered short GRBs can still be identified in GUANO searches, which employs a lower SNR detection threshold than the onboard Swift–BAT triggering criteria. The detection of S251125n exclusively in a GUANO search rather than through standard Swift–BAT triggering algorithms thus supports its identification as a genuine short GRB.

Although our discussion here assumes a fixed redshift of z=0.73, our conclusions do not depend on the specific redshift. Namely, if it were a long GRB and the flux in the observed time bin were sufficiently high, the flux in adjacent bins would not abruptly fall below the detection threshold. Therefore, even at higher redshifts, the burst would not appear as a long GRB.

\section{False Alarm Rate and Joint Probability}
This section quantifies the statistical significance of the observed spatio-temporal association between the GW signal observed by LIGO, the GRB prompt emission observed by Swift-BAT, and the X-ray afterglow observed by EP. Our objective is to determine if this alignment represents a single astrophysical source or a chance coincidence of unrelated events. 

In statistical analysis, the cornerstone of determining whether some events are independent or related is the null hypothesis, which posits that no relationship exists between the events \citep{eadie1973statistical,feller1991introduction,barlow1993statistics,feigelson2012modern}. A false alarm rate  (also known as false positive ratio) is the probability of falsely rejecting the null hypothesis. In our scenario, our goal is to quantify the probability that independent background events, instrumental ``false alarms'' or unrelated astrophysical sources, would accidentally occur with the observed proximity in time and space.

The GCN notice from LIGO for LVK S241125n gives a FAR of $9.5\times10^{-10}~{\rm Hz}$, and the Swift/BAT-GUANO sub-threshold burst carries a FAR rate of $3.7\times10^{-4}~{\rm Hz}$ \citep{2024GCN.38305....1L, 2024GCN.38308....1D}. No such value is available for the EP Follow-up X-ray Telescope (FXT) observation from GCN and published literature, so we evaluate the probability that an unrelated field source could mimic the X-ray counterpart detected during its 11 ks exposure. Since the FXT candidate was also confirmed by XRT (\url{https://gcn.nasa.gov/circulars/38324}), we therefore use the  false alarm probability (FAP) from XRT in the computation of the final joint FAR. Note that this approach would give a lower limit to the FAP of FXT, because FXT, for the same exposure time, can reach deeper flux limits than XRT and therefore could discover more spurious transients by chance. We follow the discussion on XRT rate of false positive in \citet{2016MNRAS.455.1522E}. From the LXPS page (\url{https://www.swift.ac.uk/LSXPS/docs.php}), we can check the most up\-to\-date rate at which XRT discovers new transients. To be conservative, only the total number of uncatalogued sources discovered so far (134,210) is considered, and it is divided by the total exposure time reported on the page (412\,Ms) and by the XRT field of view (0.12\,deg\textsuperscript{2}), giving an average rate of new sources found by XRT of \(R = 2.715 \times 10^{-3}\,\mathrm{s^{-1}\,deg^{-2}}\). Knowing this rate density, applying it to the area of the circle whose radius equals the angular separation between the BAT candidate and the EP candidate (i.e., \(A = 8.149 \times 10^{-3}\,\mathrm{deg}^2\)), and using the exposure time of EP \(\Delta T_{\mathrm{EP}} = 11\,\mathrm{ks}\), we estimated
\[
\mathrm{FAP}_{\mathrm{EP}} = 1 - \exp(- R \cdot A \cdot \Delta T_{\mathrm{EP}}) = 21.6\%. 
\]
A key caveat is that the BAT candidate is not a point source, as its 90\% credible region is extended. This spatial uncertainty directly impacts the EP FAP; therefore, a more comprehensive assessment should consider all EP candidates within the 90\% credible region of BAT during the 11 ks exposure. Also note that on the roles of XRT and FXT: even if both instruments detected the same candidate, this does not change the estimation of the FAP. This is because what dominates the FAP is not the probability of the event being a false positive (namely not astrophysical), but rather the probability of its being an astrophysical event not related to the GW and BAT events.

Additionally, following the RAVEN pipeline guidance (\url{https://lscsoft.docs.ligo.org/raven/joint_far.html}), we recalculate the joint FAR of the GW and BAT event, obtaining a result of $\mathrm{FAR}_{\mathrm{GW+BAT}} = 1 / 6\ \mathrm{yr}$ as reported in the GCN \url{https://gcn.nasa.gov/circulars/38356}. Then
\[
\mathrm{FAP}_{\mathrm{GW+BAT}} = 1 - \exp(-\mathrm{FAR}_{\mathrm{GW+BAT}} \cdot T_{\mathrm{obs,triple}}),
\]
where \(T_{\mathrm{obs,triple}}\) is the time during which LVK, BAT, and EP were all observing simultaneously.

This FAP should be combined with the EP FAP to get the significance for the entire event chain (GW+BAT+EP). One can write:
\[
P(A \cap B) = \mathrm{FAP}_A \times \mathrm{FAP}_{B \mid A},
\]
where $\mathrm{FAP}_{B \mid A}$ in our case would be the EP FAP, conditional on the fact that we are performing a search targeted on the GW+BAT candidate, namely 
\[
\mathrm{FAP}_{\mathrm{triple}} = \mathrm{FAP}_{\mathrm{GW+BAT}} \times \mathrm{FAP}_{\mathrm{EP}}.
\]
Finally, we can compute the FAR as
\[
\mathrm{FAR}_{\mathrm{triple}} = -\frac{1}{T_{\mathrm{obs,triple}}} \ln(1 - \mathrm{FAP}_{\mathrm{triple}}).
\]

Now, a subtle point on $T_{\mathrm{obs,triple}}$, which is the factor that makes the FAP time dependent. Note that the operations of GUANO and the full NITRATES pipeline began in $\sim$ 2022, EP officially began operations on July 22, 2024 (\url{https://ep.bao.ac.cn/ep}), and the O4 LVK runs include (\url{https://en.wikipedia.org/wiki/List_of_gravitational_wave_observations}): O4a: May 24, 2023 — January 16, 2024; O4b: April 10, 2024 — January 28, 2025; O4c: January 28, 2025 — March 31, 2025; 12 June, 2025 — November 18, 2025. One can find that the total co-observation time of the LVK runs intersected with the operational period of GUANO and the full operation of the NITRATES pipeline, together with the online period of EP, is 127 days until November 25, 2024, and is 412 days until the end of O4. We use it as the entire time during which LVK-BAT-EP observed together until the end of O4, namely totally 412 days. By substituting this value into the equation above, we obtain $\mathrm{FAP}_{\mathrm{triple}}=0.037$, which corresponds to 1.8 sigma, and $\mathrm{FAR}_{\mathrm{triple}} = 1 / 30$ yr.

\bibliographystyle{aasjournal}
\bibliography{S241125n}

\end{document}